\documentclass[12pt]{article}
\usepackage{graphicx, amsmath, amssymb, cite, setspace, color, relsize, bm, bbold, array,subfigure,verbatim}

\usepackage[top=1 in, bottom=1 in, left=.9 in, right=.9 in]{geometry}

\newcommand{\captionfonts}{\footnotesize} 
\makeatletter
\long\def\@makecaption#1#2{%
  \vskip\abovecaptionskip
  \sbox\@tempboxa{{\captionfonts #1: #2}}%
  \ifdim \wd\@tempboxa >\hsize
    {\captionfonts #1: #2\par}
  \else
    \hbox to\hsize{\hfil\box\@tempboxa\hfil}%
  \fi
  \vskip\belowcaptionskip}
\makeatother

\def\fnote#1#2{\begingroup\def\thefootnote{#1}\footnote{#2}
     \addtocounter{footnote}{-1}\endgroup}

\def\gsim{ \lower .75ex \hbox{$\sim$} \llap{\raise .27ex
\hbox{$>$}} }
\def\lsim{ \lower .75ex \hbox{$\sim$} \llap{\raise .27ex
\hbox{$<$}} }

\newcommand{\be}{\begin{equation}}
\newcommand{\ee}{\end{equation}}
\newcommand{\bea}{\begin{eqnarray}}
\newcommand{\eea}{\end{eqnarray}}
\newcommand{\beas}{\begin{eqnarray*}}
\newcommand{\eeas}{\end{eqnarray*}}

\begin{document}

\title{Flux Compactifications Grow Lumps}

\date{}

\author{Alex~Dahlen and Claire Zukowski \vspace{.1 in} \\
 \vspace{-.3 em}  
\textit{\small{Berkeley Center for Theoretical Physics, Berkeley, CA 94720, USA}}\\
 \textit{\small{\,\,Lawrence Berkeley National Laboratory, Berkeley, CA 94720, USA}}  
 }
 

\maketitle
\fnote{}{\hspace{-.65cm}emails: \tt{adahlen@berkeley.edu,czukowski@berkeley.edu}}
\vspace{-.95cm}

\maketitle

\begin{abstract}

\noindent 
The simplest flux compactifications are highly symmetric---a $q$-form flux is wrapped uniformly around an extra-dimensional $q$-sphere.  In this paper, we investigate solutions that break the internal  SO($q+1$) symmetry down to SO($q)\times\mathbb Z_2$; we find a large number of such lumpy solutions, and show that often at least one of them has \emph{lower} vacuum energy, \emph{larger} entropy, and is \emph{more} stable than the symmetric solution.  We construct the phase diagram of lumpy solutions, and  provide an interpretation in terms of an effective potential.  Finally, we provide evidence that the perturbatively stable vacua have a non-perturbative instability to spontaneously sprout lumps; we give an estimate of the decay rate and argue that generically it is exponentially faster than all other known decays.
\end{abstract}

\thispagestyle{empty} 
\newpage

\section{Introduction}

We study the ($D=p+q$)-dimensional action
\begin{gather}
S =\int d^{\,p} x\,d^{\,q}y \sqrt{-g}\left[\frac{1}{2}M_D^{D-2} \mathcal{R}-\Lambda_{D}-\frac{1}{2}\frac1{q!}\bm F_{q}^2\right]~,\label{action}
\end{gather}
where $M_D$ is the $D$-dimensional Planck mass, $\bm F_{q}$ is a $q$-form flux, and $\Lambda_{D}$ is a higher-dimensional cosmological constant.  This action admits a simple product compactification on a $q$-sphere that is uniformly wrapped by the $q$-form flux; the remaining extended $p$ dimensions form a maximally symmetric spacetime, either AdS, Minkowski, or de Sitter.   These Freund-Rubin solutions, as they are called, are considered the simplest models of stabilized extra dimensions and have a long pedigree: they were first discussed in 1980 \cite{Freund:1980xh}, and were soon generalized to non-zero $\Lambda_D$ \cite{RandjbarDaemi:1982hi}.  There has been a resurgence of interest lately because they provide a simple model for string compactifications \cite{Douglas:2006es, Silverstein:2004id,Acharya:2003ii,Denef:2008wq}.

It is known that Freund-Rubin compactifications are sometimes perturbatively unstable, and also that there exist other static extrema of Eq.~\eqref{action}:  \cite{DeWolfe:2001nz,Bousso:2002fi,Hinterbichler:2013kwa,Brown:2013mwa} showed that symmetry-breaking perturbations to the internal shape can sometimes have a negative mass squared; and~\cite{Kinoshita:2007uk,Kinoshita:2009hh,Lim:2012gh} found warped static solutions in which the internal manifold is lumpy.  
Our goal in this paper is to give a more complete story that links these observations, and to present a phase diagram of compactified solutions and their instabilities.  
In particular, we argue that:
\begin{itemize}
\item When $q\ge3$, each Freund-Rubin solution is accompanied by a large number of warped solutions where the internal manifold is lumpy;
\item Often, at least one of these lumpy solutions has \emph{lower} vacuum energy, \emph{larger} entropy, and is \emph{more} stable than the symmetric solution;
\item Perturbatively stable Freund-Rubin vacua have a previously undiscovered non-perturb-ative instability to quantum mechanically sprout lumps.  We will argue that this new decay is often the fastest  decay, proceeding exponentially faster than the two previously studied instabilities of Freund-Rubin vacua, which are flux tunneling \cite{Abbott:1984qf, BlancoPillado:2009di,Brown:2010mg,Brown:2010mf} and decompactification \cite{Giddings:2003zw, BlancoPillado:2009mi,Carroll:2009dn}.
\end{itemize}

\begin{figure}[h!] 
   \centering
   \includegraphics[width=\textwidth]{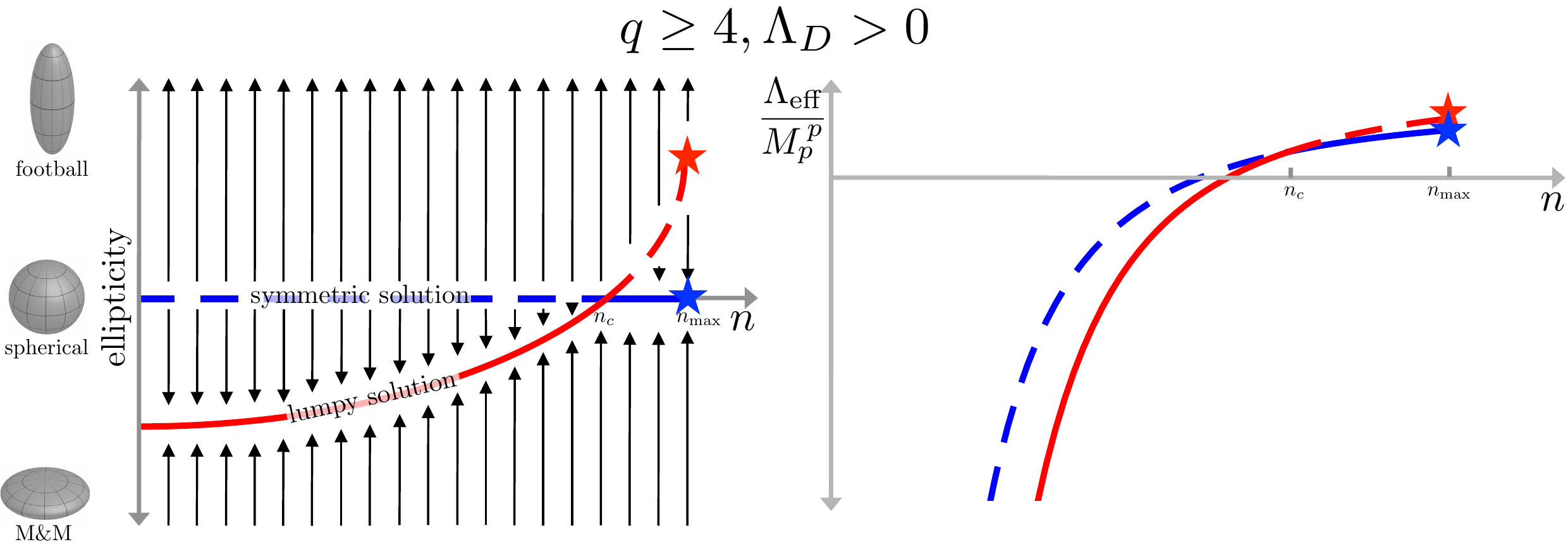}
   \caption{A cartoon of the phase digram of ellipsoidal solutions for the case $\Lambda_D>0$ and $q\ge4$.  For each value of the conserved flux number $n$, there are two solutions: a symmetric solution, where the internal manifold is a perfect sphere, and a lumpy one, where the internal manifold is either a prolate or an oblate ellipsoid.  There are two critical values of $n$: first, $n=n_c$, at which the warped solution crosses through the symmetric one; second, $n=n_\text{max}$, at which both solutions disappear spontaneously, indicated by the star.  Left: a plot of ellipticity against $n$.  Arrows indicate directions of decreasing free energy; they point towards the solution with smaller $\Lambda_\text{eff}$ and away from the solution with larger $\Lambda_\text{eff}$. 
Right: A plot of the effective cosmological constant $\Lambda_\text{eff}$ against $n$. 
The dominant solution, the one with smaller $\Lambda_\text{eff}$, is shown with a solid line and the subdominant solution is shown with a dashed line; the two solutions switch at $n=n_c$.}
\end{figure}

For example, Fig.~1 shows a sample partial phase diagram of solutions where the internal manifold is ellipsoidal.  For each value of the conserved flux number $n$, there are two solutions: the symmetric Freund-Rubin solution, and a lumpy solution where the internal manifold is deformed away from spherical.  In this case, there is a critical value $n=n_c$ at which the lumpy solution and the symmetric solution cross: for $n<n_c$, the lumpy solution is oblate (M\&M-shaped); for $n>n_c$, when the lumpy solution exists it is prolate (football-shaped).  The arrows in Fig.~1 show the directions along which the free energy is decreasing; arrows point towards the dominant vacuum and away from the subdominant vacuum.

\begin{figure}[t] 
   \centering
   \includegraphics[width= \textwidth]{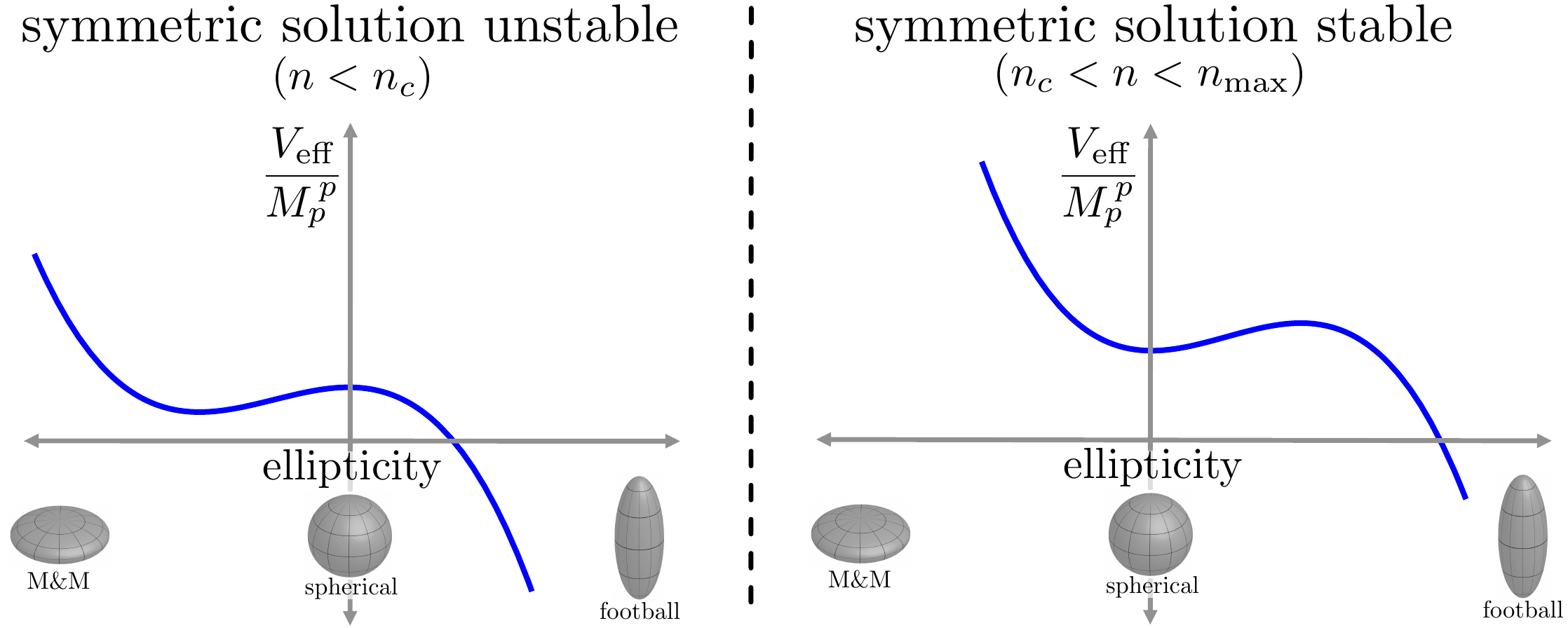}
   \caption{A cartoon of the effective potential in the ellipticity direction. We will argue that the effective potential tends to $+\infty$ as the internal manifold becomes increasingly M\&M-shaped, and to $-\infty$ as the internal manifold becomes increasingly football-shaped.  The symmetric solution is always an extremum of this effective potential.  Whether the warped solution is football- or M\&M-shaped, and whether it is dominant or subdominant is determined solely by the classical stability of the symmetric solution.  When the symmetric solution has negative mass squared (left panel), the warped solution has lower free energy and is M\&M-shaped; when the symmetric solution has positive mass squared (right panel), the warped solution has higher free energy and is football-shaped.  All these solutions have an instability (either perturbative or non-perturbative) to becoming increasingly football-shaped.}
   \label{fig:effpotcartoon}
\end{figure}

The structure of this phase diagram can be captured by the cartoon effective potential in Fig.~\ref{fig:effpotcartoon}.  In the effective potential picture, we imagine treating the ellipticity of the internal manifold as a $p$-dimensional field living in an effective potential; static solutions correspond to extrema and the value of the potential at an extremum is the effective cosmological constant $\Lambda_\text{eff}$.  The $n<n_c$ behavior is shown in the left panel: the symmetric solution has a negative mass squared; the lumpy solution is M\&M-shaped and has lower $\Lambda_\text{eff}$ than the symmetric solution.  The $n>n_c$ behavior is shown in the right panel: the symmetric solution has positive mass squared; the lumpy solution is football-shaped and has  higher $\Lambda_\text{eff}$ than the symmetric solution.  Three quantities are linked---the sign of the mass squared of the Freund-Rubin solution, whether the lumpy solution is prolate or oblate, and whether the lumpy solution is dominant or subdominant.  If you know one of these quantities, you know the other two.  In this paper, we will study different values of $\Lambda_D$, different numbers of dimensions $p$ and $q$, and even higher-$\ell$ deformations, and we will consistently find this same connection.  We argue that the essential physics of all of the lumpy compactifications is captured by the two effective potentials of Fig.~\ref{fig:effpotcartoon}.  (The effective potential is drawn for $\ell=2$ deformations, which lead to ellipsoidal solutions; we will also study higher-$\ell$ deformations, which lead  to even lumpier solutions.)

The effective potential of Fig.~\ref{fig:effpotcartoon} also implies that these solutions have a non-perturbative instability to tunnel in the football-shaped direction. 
We will provide an estimate of the rate of such a decay, and argue that it is generically the fastest non-perturbative decay of the perturbatively stable Freund-Rubin vacua.

In Sec.~2, we review the symmetric Freund-Rubin solutions and their perturbative stability.  In Secs.~3-5, we numerically construct the lumpy solutions and give the full phase diagram of solutions, first focusing on the case of the $\ell=2$ instability and the corresponding ellipsoidal solutions (Sec.~4) and then broadening our study to include higher-$\ell$ instabilities (Sec.~5).  In Sec.~6, we argue that the effective potential is that given in Fig.~\ref{fig:effpotcartoon}; we also discuss shape-mode tunneling, estimate its rate, and show that it is often the fastest known decay.  What happens to the compactification solution as it rolls down the effective potential becoming increasingly football-shaped is unclear---we speculate in the final section.  

\subsection*{Preliminaries}

The Einstein equations that follow from the action Eq.~\eqref{action} are
\begin{equation}
\label{0Einstein}
M_D^{\,D-2}\,G_{MN}=T_{MN}=\frac1{(q-1)!}F_{MP_2\cdots P_q}F_N^{\;\;\;P_2\cdots P_q}-\frac12\frac1{q!}F_q^{\;2}g_{MN}-\Lambda_{D} g_{MN}~,
\end{equation}
where $G_{MN}\equiv\mathcal{R}_{MN}-\frac12\mathcal{R}\, g_{MN}$ is the Einstein tensor, and capital Roman indices run over the full $(D=p+q)$-dimensional solution.  The Maxwell equations are
\begin{gather}
\label{0Maxwell}
(\bm{d^{\dagger}} \bm{F}_q)_{P_2\cdots P_q}=\nabla^MF_{M P_2\cdots P_q}=0~.
\end{gather}
The $q$-form flux $\bm F_q$ is the exterior derivative of a flux potential $\bm A_{q-1}$, so $\bm F_q=\bm d\bm A_{q-1}$ and $\bm d \bm F_q=0$.  

We are interested in compactified solutions where the internal manifold is topologically a $q$-sphere and the full $D$-dimensional theory is reduced down to $p$ dimensions, with $p\ge3$ and $q\ge2$.   We will use indices $\mu$, $\nu$, $\dots $ that run over the $p$ dimensions and $\alpha$, $\beta$, $\dots $ that run over the $q$ dimensions.

\section{The Symmetric Solution}
\label{sec:symmetric}

We begin by reviewing the symmetric Freund-Rubin solutions, where the $p$ extended dimensions are maximally symmetric and the $q$ internal dimensions form a $q$-sphere uniformly wrapped by flux.  The defining features of this solution follow from the fact that it is a direct product compactification.  For product compactifications, Maxwell's equations force the flux to be uniform in the extra dimensions:
\begin{gather}
\bm{F}_q= \rho \; \textbf{vol}_{S^q}~,
\end{gather}
where $\rho$ is the flux density and $ \textbf{vol}_{S^q}$ is the volume form on the internal $q$-sphere, which is proportional to the Levi-Civita tensor. 
The direct product condition also guarantees that the $p$ extended dimensions form an Einstein space. Restricting to the case of a maximally symmetric extended space-time, the metric takes the form:
\begin{gather}
ds^2=L^2 \left(-dt^2 +  \cosh^2t \, d\Omega_{p-1}^{\,2}\right)+ R^2 d\Omega_q^{\,2}~,
\end{gather}
where $L$ is the curvature length of the extended dimensions and $R$ is the radius of the internal sphere.  
The Ricci tensor is:
\begin{gather}
\mathcal R_{\mu\nu} =\frac{p-1}{L^2} g_{\mu\nu}~,\hspace{.7in} \mathcal R_{\alpha\beta} = \frac{q-1}{R^2} g_{\alpha\beta}~.
\end{gather}
When $L^{-2}>0$, the extended dimensions form a dS$_p$ with Hubble scale $H^2=L^{-2}$; when $L^{-2}<0$, analytically continuing one of the angular coordinates in the $d\Omega_{p-1}^2$ reveals that the extended dimensions form an AdS$_p$ with curvature length $\ell_\text{AdS}^{\,2}=-L^2$.  When $L^{-2}=0$, the extended dimensions are Minkowski.  

Einstein's equations, Eq.~\eqref{0Einstein}, enforce a relation between the two curvature lengths, $L$ and $R$, and the flux density, $\rho$:
\begin{gather}
\frac{\Lambda_D}{M_D^{\,D}} = \frac{(p-1)^2}2\left(M_DL\right)^{-2}+ \frac{(q-1)^2}2\left( M_D R\right)^{-2}~, \\
\frac{\rho^2}{M_D^{\,D}} = -(p-1)\left(M_DL\right)^{-2}+(q-1)\left(M_D R\right)^{-2}~.
\end{gather}

The effective $p$-dimensional cosmological constant (measured in units of the $p$-dimensional Planck mass $M_p$) is 
\begin{gather}
\frac{\Lambda_\text{eff}}{M_p^{\,p}}\equiv\frac{(p-1)(p-2)}{2}\left(M_pL\right)^{-2} = \frac{(p-1)(p-2)}{2}\left(M_DL\right)^{-2} \left(\frac{1}{M_D^{\,q}\, \text{Vol}_{S^q}}\right)^{2/(p-2)}~,
\label{LambdaEffSym}
\end{gather}
where in the last equality, we have  used the definition $M_p^{\,p-2}\equiv M_D^{\,D-2} \times \text{Vol}_{S^q}$, and $\text{Vol}_{S^q} \sim R^{^q}$ is the total internal volume.

The flux density $\rho$ is not a conserved quantity, but the total number of flux units $n$ is; $n$ is defined by integrating over the internal $q$-cycle:
\begin{gather}
n \equiv M_D^{(q-p)/2}  \int_{S_q} \bm{F}_q\,= \, \left(\frac{\rho}{M_D^{\,D/2}}\right) \,\left(M_D^{\,q} \,\text{Vol}_{S^q}\right)~.
\end{gather}
The first equality is the definition of $n$; the second is specific to the Freund-Rubin vacua.  The factor of $M_D^{(p-q)/2}$ is inserted to make $n$ dimensionless.

\begin{figure}[t!] 
   \centering
   \includegraphics[width=\textwidth]{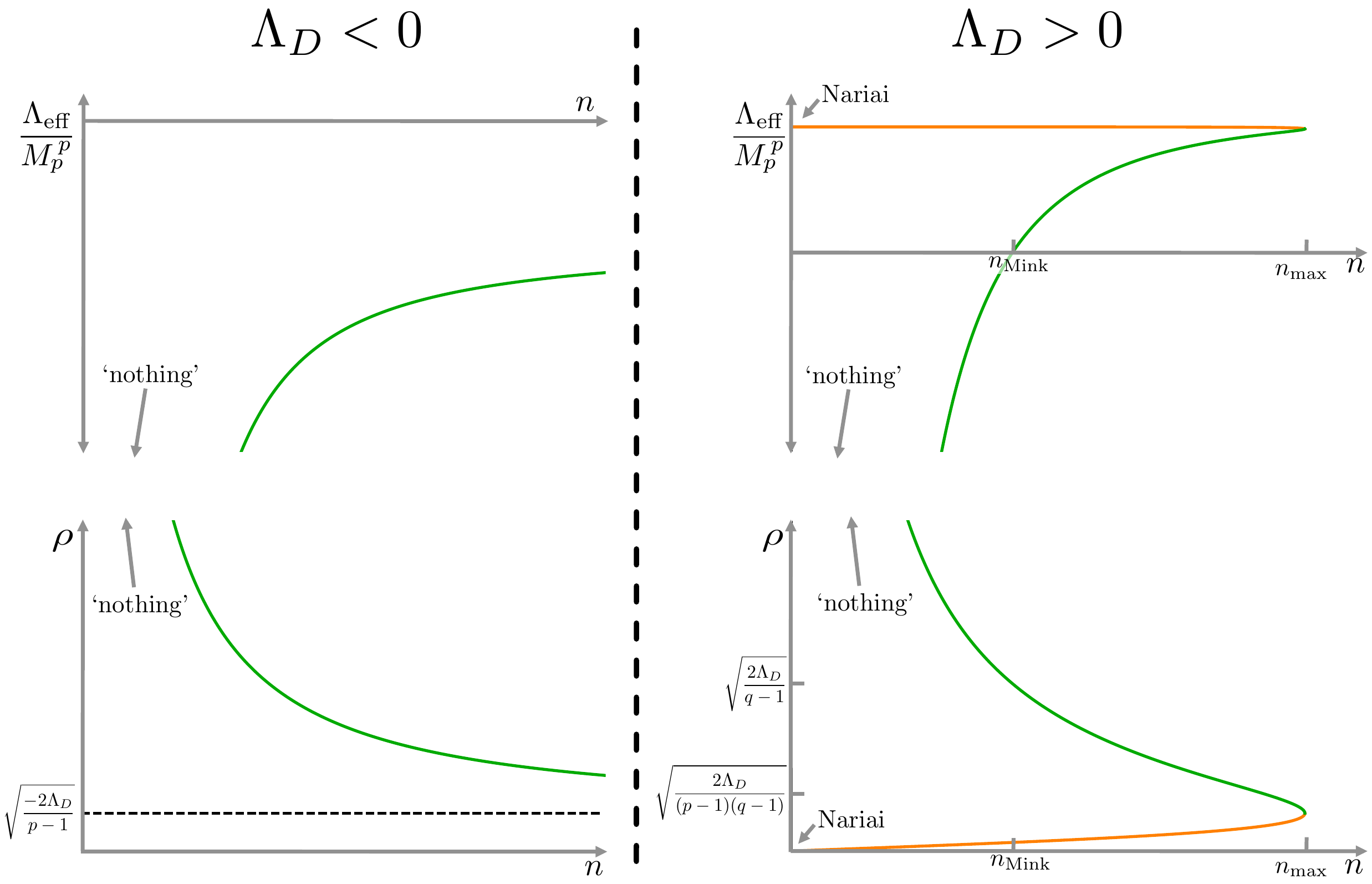}
   \caption{The Freund-Rubin solutions.  Left: When $\Lambda_D<0$, there is a single solution for each value of $n$, and it is always AdS$_p$ ($\Lambda_\text{eff}<0$).  Increasing $n$ increases $\Lambda_\text{eff}$ towards zero from below and causes the flux density $\rho$ to fall towards an asymptote. Right: When $\Lambda_D>0$, there are either two solutions or no solutions, depending on $n$.  The small-volume branch is drawn in green; it always has smaller $\Lambda_\text{eff}$, and is stable against total-volume perturbations.  It is AdS$_p$ for $n<n_\text{Mink}$ and dS$_p$ for $n_\text{Mink}<n<n_\text{max}$.  The large-volume branch is drawn in orange; it is unstable to total-volume perturbations and is always dS$_p$.  As $n$ is increased past $n_\text{max}$, the two branches merge, annihilate, and disappear.  As $n\rightarrow 0$, the green lines tend to $\Lambda_\text{eff}\rightarrow-\infty$.  The behavior of the solution in this limit is independent of $\Lambda_D$; we refer to this limit as the `nothing state'.}
   \label{fig:FRplots}
\end{figure}

For the special case $\Lambda_D=0$, these relations imply the following scalings with $n$:
\begin{gather}
M_D R \sim  n^{\frac1{q-1}}~,\hspace{.5in}\text{and}\hspace{.5in} \frac{\Lambda_\text{eff}}{M_p^{\,p}}\sim - n^{-\frac{2(D-2)}{(p-2)(q-1)}}~.
\end{gather}
For all $n$, the $p$-dimensional spacetime is AdS ($\Lambda_\text{eff}<0$).  As $n\rightarrow\infty$, $R\rightarrow\infty$ and $\Lambda_\text{eff}$ approaches 0 from below.  As $n\rightarrow0$, both the internal volume and the effective $p$-dimensional curvature length go to zero ($R\rightarrow0$ and $\Lambda_\text{eff}\rightarrow-\infty$).  We will refer to the $n\rightarrow0$ limit as the `nothing state', following the terminology of \cite{Brown:2011gt}.  The flux density $\rho\sim n R^{-q}\sim n^{-1/(q-1)}$ is inversely proportional to the number of flux units $n$.  Adding flux causes the internal radius to swell so much that the density of flux decreases.

Figure~\ref{fig:FRplots} shows the behavior of $\Lambda_\text{eff}$ and $\rho$ as a function of $n$ when $\Lambda_D\neq0$.  The behavior is qualitatively different depending on the sign of $\Lambda_D$.  When $\Lambda_D<0$, the $p$-dimensional spacetime is always AdS; increasing $n$ causes the internal volume to grow without bound and $\Lambda_\text{eff}$ to approach 0 from below.  As with the $\Lambda_D=0$ case, the flux density $\rho$ is a falling function of $n$, except that instead of asymptoting to $\rho=0$ as $n\rightarrow0$, $\rho$ approaches the nonzero value $\rho_\text{asymptote}=\sqrt{-2\Lambda_D/(p-1)}$.  There are no solutions for smaller $\rho$.  When  $\Lambda_D>0$, there is no longer only a single solution for each value of $n$.  Instead, there is a critical value $n=n_\text{max}$ at which the number of solutions changes discontinuously.  Below $n_\text{max}$ there are two solutions: a small-volume solution and a large-volume solution.  The small-volume solution always has the lower value of $\Lambda_\text{eff}$ and (as we will see in the next subsection) is always stable against total-volume fluctuations.  The large-volume solution is unstable to total-volume fluctuations: it can decrease its effective potential either by shrinking towards the small-volume branch or by expanding towards decompactification.
As $n$ is raised through $n_\text{max}$, the small-volume solution and the large-volume solution merge, annihilate, and disappear.  There are no solutions for larger $n$.

\subsection{The Effective Potential}

Another way to understand the Freund-Rubin solutions is in terms of a $p$-dimensional effective theory.  The radius $R$ is treated as a $p$-dimensional radion field, living in an effective potential given schematically by 
\begin{gather}
\frac{V_\text{eff}(R)}{M_p^{\,p}}\sim\left(\frac1{M_DR}\right)^{2q/(p-2)}\left[\frac{n^2}{(M_DR)^{2q}}-\frac1{(M_DR)^2}+\frac{\Lambda_D}{M_D^{\,D}} \right]~.
\label{effpotjustR}
\end{gather}
The three terms in square brackets represent the energy density in flux, curvature, and higher-dimensional vacuum, respectively.  The multiplicative factor outside the square brackets is related to the unit conversion from $D$-dimensional Planck units $M_D$ to $p$-dimensional Planck units $M_p$---the same factor that appeared in Eq.~\eqref{LambdaEffSym}.  The flux term dominates at small $R$; flux lines repel and push the sphere out to larger radius.  The curvature is an attractive term and the two terms can interact to form a minimum of the potential.  The Freund-Rubin solutions  are solutions in which the scalar field remains static at an extremum of this effective potential, and the value of the potential at that extremum is $\Lambda_\text{eff}$.  

\begin{figure}[t!] 
   \centering
   \includegraphics[width=\textwidth]{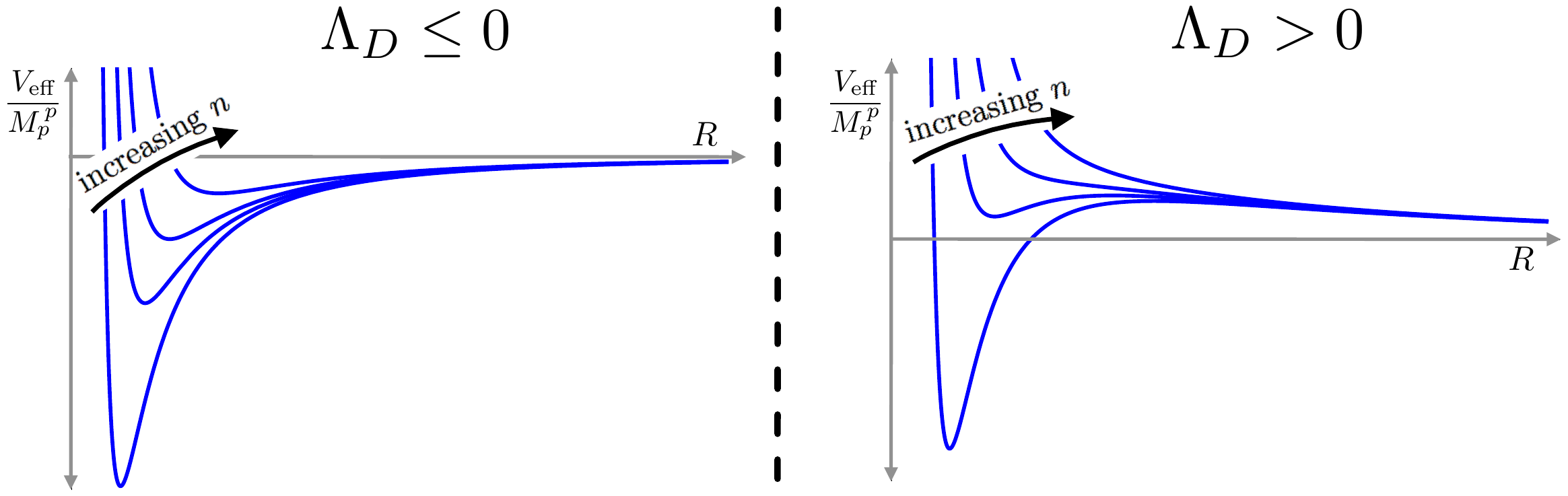}
   \caption{The effective potential for the radion field $R(x)$ using an ansatz that assumes spherical symmetry of the internal dimensions.  The extrema of this potential correspond to the Freund-Rubin solutions discussed in the previous section.  Left: When $\Lambda_D\le0$, there is a single minimum for each value of $n$; increasing $n$ causes the solution to move out to larger $R$ and less negative $\Lambda_\text{eff}/M_p^{\,p}$.  Right: When $\Lambda_D>0$, the number of extrema depends on $n$.  For small $n$, the effective potential has a minimum and a maximum.  The maximum is always de Sitter, and the minimum can be either de Sitter or AdS.  Increasing $n$ causes the minimum and maximum to merge and annihilate.}
   \label{fig:FReffpot}
\end{figure}

This effective potential is plotted in Fig.~\ref{fig:FReffpot} for various values of $n$. The qualitative behavior depends on the sign of $\Lambda_D$.  When $\Lambda_D\le0$, there is only ever a single extremum, which is always an AdS minimum; increasing $n$ shifts the minimum to larger values of $V_\text{eff}$ and to larger values of $R$, in agreement with the results of Fig.~3.  When $\Lambda_D>0$, there are two extrema---a minimum and a maximum which come together, merge, and annihilate as $n$ is increased through $n_\text{max}$.  

At small $n$ (specifically $n\ll(\Lambda_D/M_D^D)^{\,(q-1)/2}$), the behavior of the minimum becomes independent of $\Lambda_D$.  This is because, at such small $n$, the minimum sits at a value of $R$ that is hierarchically smaller than the higher-dimensional Hubble length $H_D^{-1}\sim M_D (\sqrt{\Lambda_D/M_D^{\,D}})^{-1}$.  This hierarchy means that higher-dimensional curvature cannot affect the compactification.

To arrive at this effective potential, we treated the shape of the internal sphere as fixed, and the radius of the sphere as a dynamic field.  This assumption, however, proves too restrictive: minima of the effective potential in Fig.~4 can be unstable saddle points in additional directions in field-space that correspond to shape-mode fluctuations.

\subsection{Stability}
The full perturbative spectrum of these Freund-Rubin solutions was computed in \cite{DeWolfe:2001nz,Bousso:2002fi,Hinterbichler:2013kwa,Brown:2013mwa}, and some diagonalized fluctuations were shown to have a negative mass squared.  In this subsection, we will review the relevant parts of the calculation (deferring difficulties to those papers) and extract some information that will be relevant for us.
In particular, we will leave all fluctuations turned off except for two scalar perturbations with angular momentum $\ell\ge2$: a shape mode in which the internal manifold deforms away from sphericality and a flux mode in which the flux distribution deforms away from uniformity; that these two modes decouple from all other fluctuations is proven in \cite{DeWolfe:2001nz,Bousso:2002fi,Hinterbichler:2013kwa,Brown:2013mwa}.

The shape mode is a fluctuation of the metric of the form:
\begin{gather}
 \delta g_{\mu\nu} = -\frac1{p-2} g_{\mu\nu}h(x) Y_\ell (\theta)~,\hspace{.5in}\delta g_{\alpha\beta}=\frac1q g_{\alpha\beta} h(x) Y_\ell (\theta)~.
\end{gather}
The second half of this equation says the internal sphere is deformed along the $Y_\ell$ spherical harmonic by an amount proportional to the $p$-dimensional field $h(x)$.  The first half says that the $p$-dimensional metric $g_{\mu\nu}$ adjusts by $-hY_\ell/(p-2)$, introducing warping into the compactification; this adjustment is required to decouple the fluctuation from the $p$-dimensional graviton---it is the linearized version of the Weyl transform that changes units from $M_D$ to $M_p$.  The scalar flux mode is a perturbation to the gauge field of the form:
\begin{gather}
\delta A_{\alpha_1\cdots\alpha_{q-1}}=M_D^{-2} a(x) \rho \,  \epsilon^{\beta}_{\,\,\,\alpha_1\cdots\alpha_{q-1}}\nabla_\beta Y_\ell(\theta)~.
\end{gather}
This corresponds to a change in the field strength tensor:
\begin{gather}
\delta  F_{\alpha_1\cdots\alpha_q}= - a(x)  \rho \,\epsilon_{\alpha_1\cdots\alpha_q}\,\lambda_\ell Y_\ell(\theta)~,\hspace{.3in}\delta  F_{\mu \alpha_2\cdots\alpha_q}=M_D^{-2}\nabla_\mu a(x) \rho\, \epsilon^\beta_{\,\,\,\alpha_2\cdots\alpha_q}\nabla_\beta  Y_\ell(\theta)~,
\end{gather}
where $\lambda_\ell=\ell(\ell+q-1)/(M_D^{\,2}R^2)>0$ is the eigenvalue of the spherical harmonic, $\Box_y Y_\ell(\theta)=-\lambda_\ell Y_\ell(\theta)$; under this perturbation, the flux density around the internal manifold shifts along the $-Y_\ell$ spherical harmonic by an amount proportional to the $p$-dimensional field $a(x)$.  There is also an adjustment to the component of $\bm F_q$ with one index off the sphere; this adjustment is necessary to enforce Bianchi's identity.
Both $h$ and $a$ are dimensionless as defined.

These two fluctuations decouple from all other fluctuation modes; they can be diagonalized to form two linearly independent fluctuations $\psi_\pm$ which satisfy 
\begin{gather}
\Box_x \psi_\pm=m_\pm^{\,2}\psi_\pm~,\label{waveeqn}
\end{gather}
with
\begin{gather}
\psi_\pm=\left(A\pm\sqrt{A^2+4\lambda_\ell B}\right)h +2\frac{q-1}q \rho^2 a~,
\hspace{.3in}
\frac{m_\pm^{\,2}}{M_D^{\,2}}=A+\lambda_\ell \pm \sqrt{A^2+4\lambda_\ell B}~,\label{eigenvecvals}
\end{gather}
and
\begin{gather}
A= \frac{q(p-1)}{D-2}\frac{\rho^2}{M_D^{\,D}}-\frac{q-1}{M_D^{\,2}R^2}~,\hspace{.3in} B=\frac{(p-1)(q-1)}{D-2}\frac{\rho^2}{M_D^{\,D}}~.\label{AandB}
\end{gather}

Equations~\eqref{waveeqn}-\eqref{AandB} are only valid for angular momenta $\ell\ge2$.  This is because the $\ell=0$ flux mode is gauge (adding a constant to $\bm A_{q-1}$ does not change $\bm F_q$) and the $\ell=1$ shape mode is gauge (perturbing a sphere by its $\ell=1$ harmonic shifts it but does not change the induced metric on it).  For $\ell\ge2$, however, neither mode is gauge, so both $\psi_+$ and $\psi_-$ are physical fluctuations.  The mode $\psi_+$ always has a positive mass squared; $\psi_-$ on the other hand is the danger mode, as it can sometimes have a negative mass squared.  For this danger mode, $h$ and $a$ shift in opposite directions ($\text{sign } h = -\text{sign } a$); this means that wherever the radius gets larger, the flux density also gets larger, and vice versa ($\text{sign }\delta F_{\alpha_1\dots\alpha_q}=\text{sign }\delta g_{\alpha\beta}$).  For instance, in the unstable $\ell=2$ direction, when the internal manifold becomes football-shaped, the flux concentrates at the poles, and when the internal manifold becomes M\&M-shaped, the flux concentrates at the equator.  The $\psi_+$ mode, the safe mode, has the opposite behavior.

The $\ell$th danger mode has a negative mass squared ($m_-^{\,2}<0$) if and only if
\begin{gather}
\frac{\rho^2 R^2}{M_D^{D-2}}=\frac{1}{p-1}\left[(q-1)(D-2)-2 \frac{\Lambda_D R^2}{M_D^{D-2}}\right]> \frac{D-2}{2(p-1)(q-2)}\bigg[\ell(\ell+q-1)-2(q-1)\bigg]~.
\label{cRcritical}
\end{gather}

The implications of Eq.~\eqref{cRcritical} are plotted in Fig.~5, and can be summarized as follows:
\begin{itemize}
\item When $\Lambda_D=0$, the $\ell=(q-1)$ mode is exactly massless; all modes between $\ell=2$ and $\ell=(q-2)$ have negative mass squared and all higher-$\ell$ modes have positive mass squared.  
\item When $\Lambda_D\neq0$, the $n\rightarrow0$ limit has the same stability properties as the $\Lambda_D=0$ case.  In this limit (which we label as `nothing' in Fig.~5) $R\rightarrow0$, $\rho\rightarrow\infty$, and the value of $\rho^2R^2$ in Eq.~\eqref{cRcritical} becomes independent of $\Lambda_D$.  
\item When $\Lambda_D>0$, there are fewer $\ell\ge2$ modes  that have a negative mass squared, and increasing $R$ tends to make  more modes stable.  The `Nariai' solution, with $\rho=n=0$, is unstable to the total-volume mode (which has $\ell=0$) but stable to all modes with $\ell\ge2$.
\item When $\Lambda_D<0$, there are more $\ell\ge2$ modes that have a negative mass squared, and increasing $R$ tends to make  more modes unstable.  As $n$ and $R$ go to infinity, eventually all danger modes will develop a negative mass squared.
\end{itemize}

\begin{figure}[b!] 
   \centering
   \includegraphics[width=\textwidth]{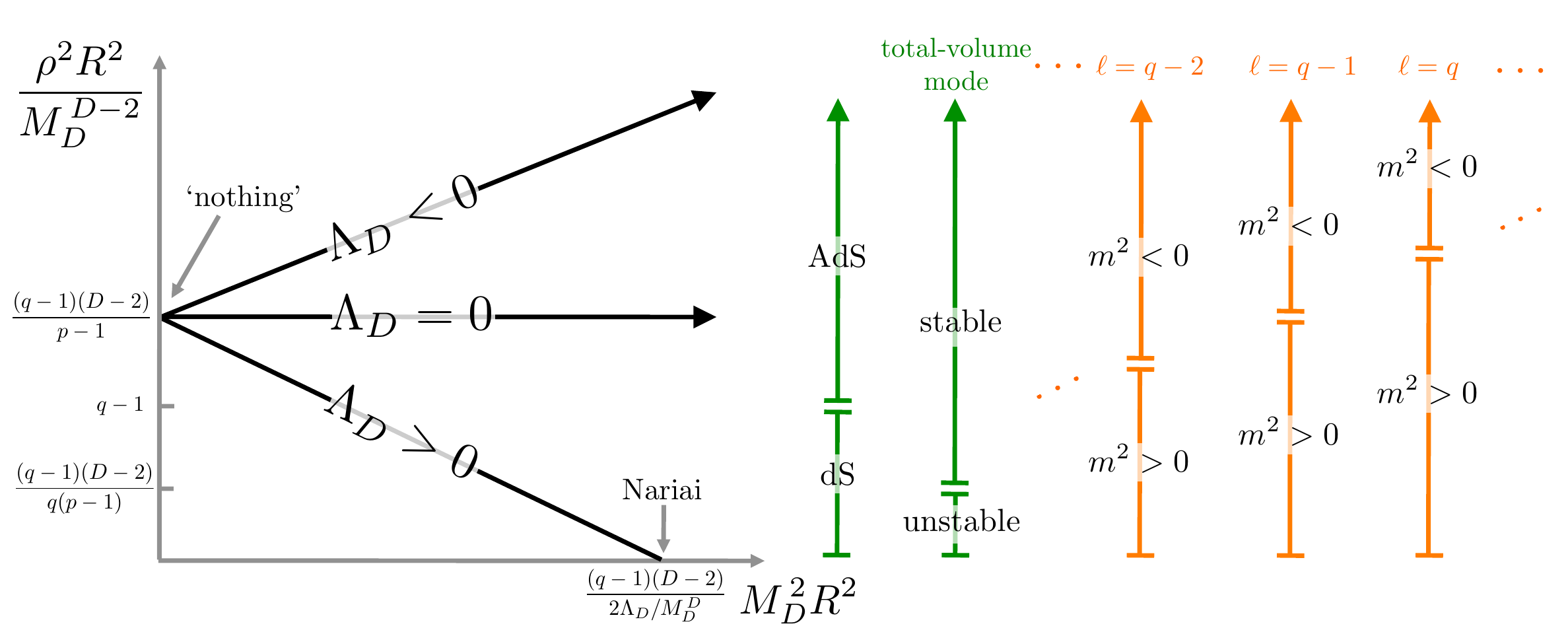}
   \caption{Equation~\eqref{cRcritical} gives the critical value of $\rho^2R^2$ at which the $\ell$th spherical harmonic develops a negative mass squared.  The larger $\ell$ is, the larger the value of $\rho^2R^2$ at which the mode becomes negative.  When $\Lambda_D=0$, there is a fluctuation with $\ell=q-1$ that is exactly massless, and there are fluctuations with $\ell=2$ through $\ell=q-2$ that have negative mass squareds.  When $\Lambda_D>0$, increasing $R$ tends to push more modes to positive mass squareds.  When $\Lambda_D<0$, increasing $R$ tends to push more modes to negative mass squareds; in fact, as $R\rightarrow\infty$, there are negative mass squared modes for all values of $\ell$.}
\end{figure}

For example, let us look in detail at the case $p=q=4$.  When $\Lambda_D=0$, the $\ell=2$ mode has a negative mass squared, the $\ell=3$ mode is perfectly massless, and all higher modes are massive.  
When $\Lambda_D<0$, increasing $n$ makes higher and higher $\ell$ modes develop a negative mass squared. 
When $\Lambda_D>0$, there are two critical values of $n$.  First, there is the value $n=n_\text{max}\equiv81\pi^2/[\sqrt{2}(\Lambda_8/M_8^{\,8})^{3/2}]$; at this value the small-volume and large-volume branches of the Freund-Rubin solutions merge and annihilate.  Second, there is the value $n=n_c\equiv32\sqrt{3}\pi^2/(\Lambda_8/M_8^{\,8})^{3/2}\sim.97\,n_\text{max}$; at this value, the $\ell=2$ fluctuation about the small-volume branch is perfectly massless.  For all $n<n_\text{max}$, the large-volume branch is unstable to the total-volume mode, but stable to all higher-$\ell$ modes.  The small-volume branch is always stable to the total-volume mode and to all modes with $\ell\ge3$, but the mass squared of the $\ell=2$ danger mode changes at $n=n_c$.  When $n<n_c$, the danger mode has a negative mass squared, and when $n_c<n<n_\text{max}$, the danger mode has a positive mass squared.  Only the small-volume solutions with $n_c<n<n_\text{max}$ are completely perturbatively stable.

(AdS compactifications can tolerate modes with small negative mass squareds and still remain stable, as long as the mass squared is above the BF bound \cite{Breitenlohner:1982bm}.  The results summarized above concern where the modes develop a negative mass squared, not where they go unstable.  The critical value where $m^2=0$ is the more important one for this paper; for results on stability, see  \cite{DeWolfe:2001nz,Bousso:2002fi,Hinterbichler:2013kwa,Brown:2013mwa}.)

\section{The Lumpy Solutions}
\label{sec:lumpy}

In the previous section, we investigated the Freund-Rubin solutions, where the internal manifold is a $q$-sphere uniformly wrapped by $q$-form flux;   
we saw that symmetry-breaking perturbations can sometimes have a negative mass squared.  The fact that there is a critical value of $n$ at which a perturbation develops a negative mass squared implies that there must be other lumpy solutions to the same equations of motion.  (Under smooth deformations that preserve the asymptotics, a local minimum of a one-dimensional function cannot become a local maximum without ejecting other extrema.)  The goal of this paper is to construct these warped, lumpy solutions and to understand their properties.  

These additional solutions do not have perfect symmetry and uniformity, and therefore they necessarily include warping.  
For simplicity, we investigate only solutions that are warped along a single internal direction---in other words, solutions that break the SO($q+1$) symmetry of the Freund-Rubin compactifications down to an SO($q$) symmetry.  We additionally assume that the internal manifold has a symmetry under exchange of the north and south poles, so the full internal symmetry will be SO($q)\times\mathbb Z_2$.  Our symmetry ansatz means that we study spherical harmonics $Y_\ell$ with $\ell$ even and zero azimuthal part; a more complete study would further break this internal symmetry, but even with this restrictive ansatz, we still find a large number of lumpy solutions.  

Our metric ansatz is:
\begin{gather}
 ds^2= \Phi(\theta)^2\left[-dt^2+\cosh^2 t\, d\Omega_{p-1}^2\right]+R(\theta)^2 d\Omega_q^{\,2}~,\label{warpedmetric}
 \end{gather}
 where $\theta$ is the angular direction singled out for warping, and $d\Omega_q^{\,2}=d\theta^{\,2}+\sin^2\theta d\Omega_{q-1}^{\,2}.$  The flux is also taken to be non-uniformly distributed in the $\theta$-direction:
 \begin{gather}
F_{\alpha_1\cdots \alpha_q}= M_D^{(q-p)/2} Q \Phi^{-p}(\theta) \epsilon_{\alpha_1\cdots \alpha_q}~,
\label{warpedflux}
\end{gather}
where $Q$ is a constant, and the factor of $M_D^{(q-p)/2}$ is included to make $Q$ dimensionless. This flux ansatz automatically satisfies both Maxwell's equations Eq.~\eqref{0Maxwell} and Bianchi's identity $\bm d \bm F_q=0$.

Plugging Eqs.~\eqref{warpedmetric} and \eqref{warpedflux} into Eq.~\eqref{0Einstein} gives one  constraint equation,
\footnotesize
\be \frac{2\Lambda_D R^2}{M_D^{D-2}} = \left(\frac{Q R}{M_D^{p-1}\Phi_{\textcolor{white}{a}}^{p}}\right)^2+p(p-1)\frac{R^2-\Phi'^2}{\Phi^2} - 2p(q-1)\frac{R\cot{\theta}+R'}{R}\frac{\Phi'}{\Phi}+ (q-1)(q-2) \frac{R^2-2RR'\cot\theta-R'^2}{R^2}~,\\ \label{constraint} \ee
\normalsize
and two dynamic equations of motion,
\footnotesize
\begin{align}
\hspace{-.5in}
(D-2)\frac{R''}{R} &= -p\left(\frac{Q R}{M_D^{p-1}\Phi_{\textcolor{white}{b}}^{p}}\right)^2-p(p-1)\frac{R^2-\Phi'^2}{\Phi^2}+p(q-p)\frac{R\cot{\theta}+R'}{R}\frac{\Phi'}{\Phi}+(D-2)\frac{R^2-RR'\cot{\theta}+R'^2}{R^2}\nonumber \\
& \hspace{0.35cm} + (p-1)(q-2)\frac{R^2-2RR'\cot{\theta}-R'^2}{R^2}~,\label{eqn1}\\
(D-2)\frac{\Phi''}{\Phi}&=(q-2)\left(\frac{Q R}{M_D^{p-1}\Phi_{\textcolor{white}{c}}^{p}}\right)^2+(p-1)(q-2)\frac{R^2-\Phi'^2}{\Phi^2}+ (q-1)(p-q+2)\frac{R\cot{\theta}+R'}{R}\frac{\Phi'}{\Phi}+(D-2)\frac{R' \Phi'}{R\Phi}\nonumber \\
&\hspace{0.35cm} -(q-1)(q-2)\frac{R^2-2RR'\cot{\theta}-R'^2}{R^2}~.\label{eqn2}
\end{align}
\normalsize
These equations admit a constant solution where $R=R_0$ and $\Phi=\Phi_0$;  this is the Freund-Rubin solution of Sec.~2 with $L=\Phi_0$ and $\rho=M_D^{(q-p)/2} QL^{-p}$.  But constants are not the only solutions.

\begin{figure}[h!] 
   \centering
   \includegraphics[width=.84\textwidth]{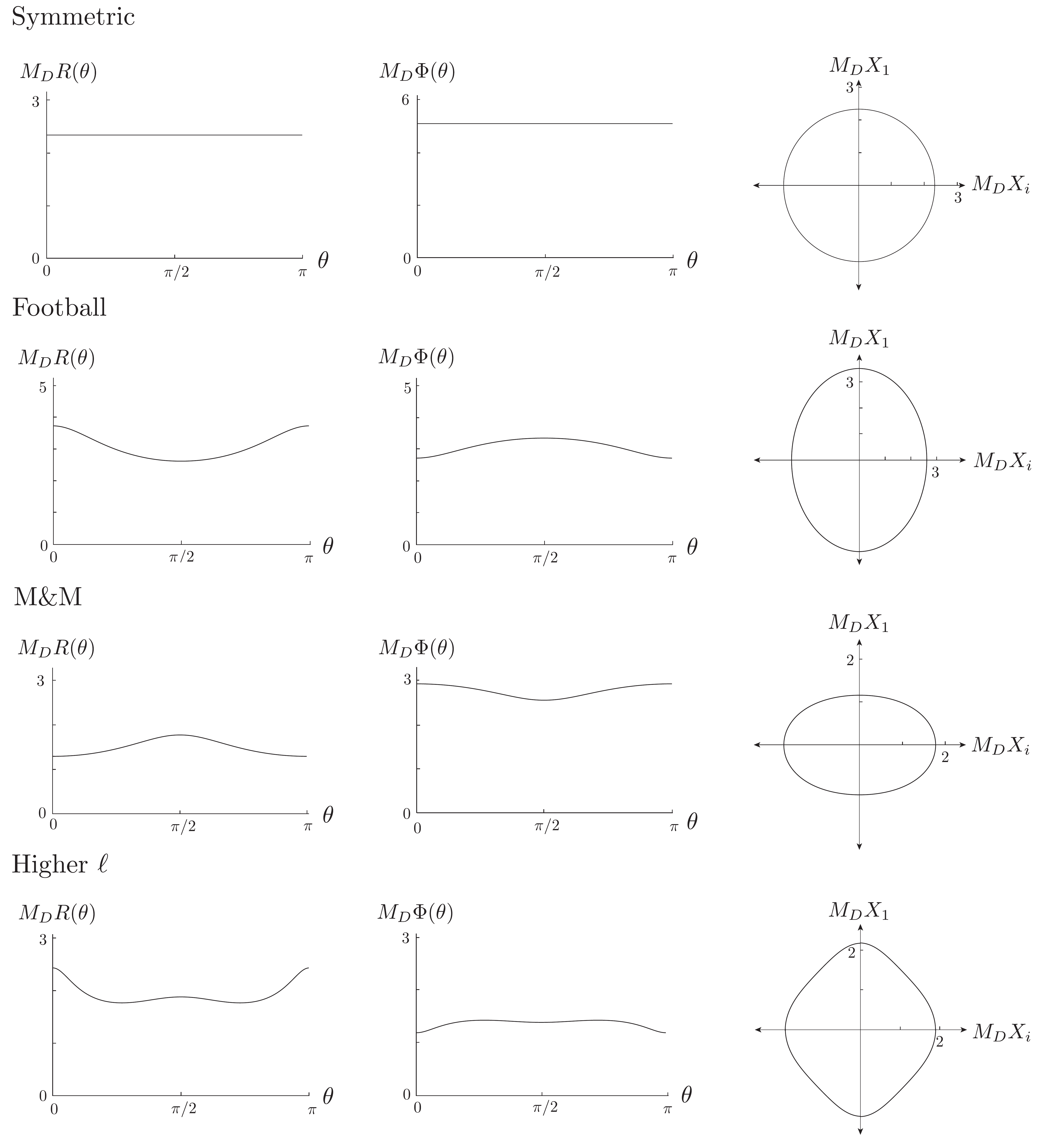}
   \caption{Four sample solutions to Eqs.~\eqref{constraint}-\eqref{eqn2}, all with $p=q=4$.  The top solution, labeled `symmetric' [$\Lambda_D/M_D^{\,D}=1$, $Q=446$ and $M_D\Phi(\pi/2)=5.10$] is of the Freund-Rubin type, where $R$ and $\Phi$ are constant. The next two solutions, labeled `football' [$\Lambda_D/M_D^{\,D}=1$, $Q=34.4$ and $M_D\Phi(\pi/2)= 3.36$] and `M\&M' [$\Lambda_D/M_D^{\,D}=1$, $Q=58.6$ and $M_D\Phi(\pi/2)=2.55$], both have a single extremum between $\theta=0$ and $\theta=\pi$. These solutions are ellipsoidal in shape and related to the $\ell=2$ instability.      The last solution, labeled `Higher $\ell$' [$\Lambda_D/M_D^{\,D}=-1$, $Q=5.83$ and $M_D\Phi(\pi/2)=1.37$] has three extrema between $\theta=0$ and $\theta=\pi$.
   There can be many solutions with the same value of $n$---the top three solutions all come from the same theory, and have $n=.91n_\text{max}$.  The symmetric solution lies on the small-volume Freund-Rubin branch of Fig.~\ref{fig:8Ddsphasediagram}, the M\&M lies on the small-volume lumpy branch, and the football lies on the large-volume lumpy branch.
    The third column gives the internal geometry as an embedding space: when the plotted curve is rotated around the $X_1$-axis, the induced metric on the resulting surface is equal to the  metric of the lumpy internal manifold.  Football solutions are prolate and $R$ reaches a minimum at the equator ($\theta=\pi/2$); M\&M solutions are oblate and $R$ reaches a maximum at the equator.  The field $\Phi$ always has the opposite behavior, so that the flux density always gets larger where $R$ gets smaller, and vice versa.}
   \label{fig:samplesols}
\end{figure}

We find non-constant solutions numerically; such lumpy solutions were first found in \cite{Kinoshita:2007uk}, and sample solutions that we found are shown in Fig.~\ref{fig:samplesols}. 
The number of extrema that a solution has between $\theta=0$ and $\theta=\pi$ will turn out to be an important classifier.  In Sec.~\ref{ell2}, we will study solutions with one extremum, such as the middle two lines of Fig.~\ref{fig:samplesols}; these solutions are ellipsoidal in shape and will turn out to be related to  $\ell=2$ deformations of the Freund-Rubin solutions.  In Sec.~\ref{higherell}, we will study solutions with even more extrema, such as the bottom line of Fig.~\ref{fig:samplesols}; these solutions are even lumpier than ellipsoids and will turn out to be related to higher-$\ell$ harmonics.  The remainder of this section is devoted to a discussion of a few properties of the equations of motion, Eqs.~\eqref{constraint}-\eqref{eqn2}.

\vspace{.1in}

\noindent \textbf{The Case $\bm{q=2}$:}
When the internal manifold is topologically a 2-sphere, the constraint equation enforces that the Freund-Rubin solution is the only solution to the equations of motion.  There are no lumpy solutions that fit our metric ansatz unless $q\ge3$. \vspace{.1in}

\noindent \textbf{Scaling Properties:}
Equations~\eqref{constraint}-\eqref{eqn2} have a scaling symmetry that relates solutions at different values of $\Lambda_D$.    The equations of motion are invariant under the transformation
\begin{gather}
\Lambda_D\rightarrow \alpha^{-2} \Lambda_D~, \hspace{.4in}  R(\theta)\rightarrow \alpha R(\theta)~, \hspace{.4in} \Phi(\theta)\rightarrow\alpha \Phi(\theta)~, \hspace{.4in} Q\rightarrow\alpha^{(p-1)}Q~.
\end{gather}
This means that there is no need to study different values of $\Lambda_D$, because the physics is simply related by scaling.  The one restriction on this argument is that $\alpha^2$ is necessarily positive, so this scaling transformation does not map positive $\Lambda_D$ to negative $\Lambda_D$. 
There are thus three cases to consider: $\Lambda_D<0$,  $\Lambda_D=0$, and $\Lambda_D>0$. 

When $\Lambda_D=0$, the equations have an additional scaling symmetry that relates solutions at different values of $Q$.  The $\Lambda_D=0$ version of the equations of motion are invariant under the transformation
\begin{gather}
Q\rightarrow\beta^{p-1} Q~, \indent \Phi(\theta)\rightarrow \beta \Phi(\theta)~, \indent R(\theta)\rightarrow \beta R(\theta) \indent (\text{when }\Lambda_D=0)~.
\end{gather}
In this case, we need only consider a single value of $Q$, because solutions with different amounts of flux are related by rescaling the field profiles $R(\theta)$ and $\Phi(\theta)$. 
In other words, when $\Lambda_D=0$, adding more flux increases the total internal volume, but does not affect the shape or flux distribution of the solution.  \vspace{.1in}

\noindent \textbf{Boundary Conditions:}
Regularity at the north and south poles demands that $R'(0)=\Phi'(0)=R'(\pi)=\Phi'(\pi)=0$; these conditions are also implied by the equations of motion at the poles. 
As discussed, we also assume the internal manifold has a $\mathbb Z_2$ symmetry that relates the north and south poles; this extra symmetry implies that both $R$ and $\Phi$ are even functions about the equator $\theta=\pi/2$. 

To find solutions, we use a shooting technique.  We set initial conditions for $R$ and $\Phi$ at the equator, and numerically evolve to one of the poles.  For most initial conditions, regularity is not satisfied at the poles, but there is a measure zero set of initial conditions at $\theta=\pi/2$ that gives the appropriate behavior at $\theta=0$ and $\pi$: these are our solutions.  Initial conditions at the equator are $R(\pi/2)$ and $\Phi(\pi/2)$; the constraint equation solves for $R(\pi/2)$, so for each value of $Q$ and $\Lambda_D$, we scan over values of $\Phi(\pi/2)$ to find solutions---there can be several for each set of $Q$ and $\Lambda_D$. \vspace{.1in}

\noindent \textbf{Computing $\bm{\Lambda_\text{eff}}$:}
To compute $\Lambda_\text{eff}/M_p^{\,p}$ for a numeric solution, we use the action density.  Numeric solutions are plugged into the action Eq.~\eqref{action}, which is then integrated over the internal dimensions.  The result is compared against the formula for a $p$-dimensional maximally symmetric spacetime with cosmological constant $\Lambda_\text{eff}$ and Planck mass $M_p$:
\begin{gather}
S_\text{E}=\int d^px \, M_p^{\,p} \sqrt{g_p} \,\frac2{p-2}\left(\frac{\Lambda_\text{eff}}{M_p^{\,p}}\right)~.
\end{gather}
For the symmetric solutions, this definition agrees with the formula Eq.~\eqref{LambdaEffSym}.   The smaller $\Lambda_\text{eff}/M_p^{\,p}$, the smaller the free energy; the solution with the lowest value of $\Lambda_\text{eff}$ at a given value of the conserved flux number $n$ is the dominant vacuum. \vspace{.1in}

\noindent \textbf{Embedding:}
As a visualization tool, we will sometimes plot the internal metric as an embedding in $q+1$ Euclidean dimensions; the right column of Fig.~\ref{fig:samplesols} gives such an embedding.  The internal metric $ds^2=R(\theta)^2d\Omega_q^{\,2}$ can be realized as the induced metric on the surface
\begin{eqnarray}
X_1&=&\int_{\pi/2}^{\theta} d\theta' \sqrt{R(\theta')^2-[\partial_{\theta'}(R(\theta')\sin\theta')]^2}\\
X_2&=&R(\theta) \sin\theta\cos\theta_2\\
X_3&=&R(\theta)\sin\theta\sin\theta_2\cos\theta_3\\
&\vdots&\nonumber \\
X_q&=&R(\theta)\sin\theta\sin\theta_2\cdots\sin\theta_{q-1}\cos\theta_q\\
X_{q+1}&=&R(\theta)\sin\theta\sin\theta_2\cdots\sin\theta_{q-1}\sin\theta_q
\end{eqnarray}
in $(q+1)$-dimensional flat space with coordinates $(X_1,\dots,X_{q+1})$.   
Where available, we will give embedding surfaces because they provide an intuitive way to picture the internal geometry.
When $R(\theta)^2<[\partial_{\theta}(R(\theta)\sin\theta)]^2$, i.e.~when $R(\theta)$ is far from constant, the embedding surface is no longer real; instead it pivots into the complex $X_1$-plane.  In this case, analytically continuing $X_1\rightarrow iX_1$ realizes the metric as an embedding in $(q+1)$-dimensional Minkowski space, though perhaps this final step belies any gains in intuition.

\section{Ellipsoidal Solutions and the $\bm{\ell=2}$ Instability}
\label{ell2}
We first examine the ellipsoidal solutions, which have a single extremum between $\theta=0$ and $\theta=\pi$.  The middle two rows in Fig.~\ref{fig:samplesols} are examples of ellipsoids, and we will argue that these solutions are related to the $\ell=2$ perturbations of the Freund-Rubin solution.  

The first clue that they are related to the $\ell=2$ instability comes from the shape of these single-extremum solutions.  Not only is the field profile $R(\theta)$  roughly ellipsoidal, but the flux is distributed on the ellipsoid in the appropriate way.  
 We saw in Sec.~2.2 that the unstable danger mode had shape mode $h$ and flux mode $a$ inversely correlated; likewise, all of the solutions we find have $R$ and $\Phi$ inversely correlated, as can be seen in Fig.~\ref{fig:samplesols}.  The flux density gets larger wherever the radius of the sphere gets larger, so that the flux is concentrating at the tips of the footballs and around the equator of the M\&Ms.  This makes intuitive sense because, for these solutions, the regions with larger radius have higher curvature and it takes a larger flux density to support a region of higher curvature against collapse.

\begin{figure}[t] 
   \centering
   \includegraphics[width=.7\textwidth]{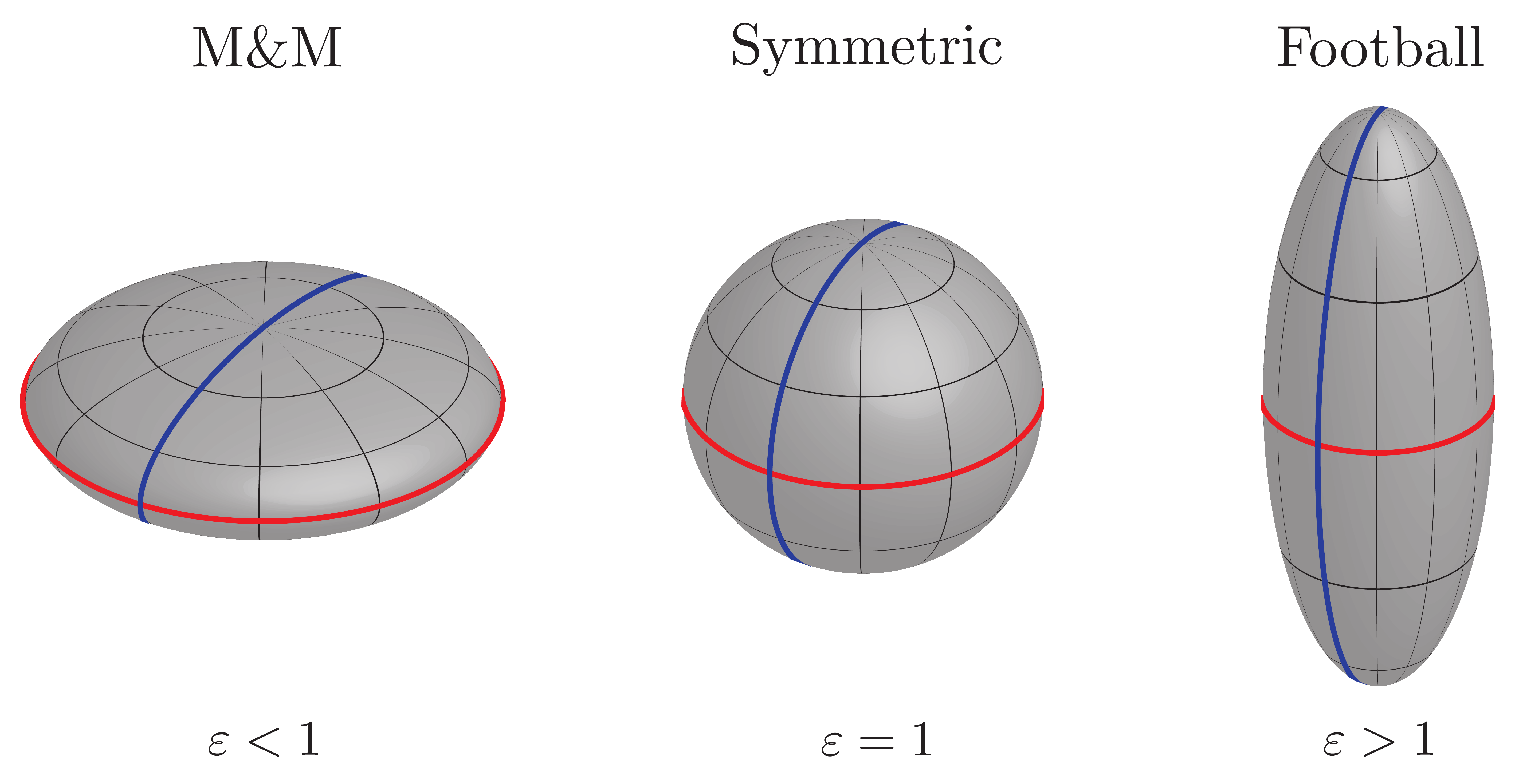}
   \caption{The order parameter $\varepsilon$, which measures the lumpiness of the solution, is defined by the length of the blue curve that runs along a line of longitude divided by the length of the red curve that runs around the equator.  M\&M-shaped solutions have $\varepsilon<1$; football-shaped solutions have $\varepsilon>1$; Freund-Rubin solutions have $\varepsilon=1$ by definition.}
      \label{fig:ellipticitydef}
\end{figure}

For each of these lumpy solutions, we define an order parameter that quantifies the ellipticity:
\begin{gather}
\varepsilon\equiv\frac{2\times(\text{distance from north to south pole})}{\text{distance around equator}}=\frac{2\int_0^\pi R(\theta)d\theta}{2\pi R(\theta=\pi/2)}~.
\end{gather}
Figure~\ref{fig:ellipticitydef} shows the definition of $\varepsilon$ pictorially.  Football-shaped solutions have $\varepsilon>1$ and M\&M-shaped solutions have $\varepsilon<1$.  (We opted for this definition of ellipticity because it is intrinsic to the internal manifold, rather than the standard definition of ellipticity which appeals to embedding space.)

In the remainder of this section, we will present phase diagrams of ellipsoidal solutions.
We will see that there are three simple principles that determine the shape of the phase diagrams:
\begin{itemize}
\item Every Freund-Rubin solution is accompanied by a single ellipsoidal solution with the same value of $n$.
\item Whenever the $\ell=2$ `danger mode' has a negative mass squared, the ellipsoidal solution is M\&M-shaped and energetically favored [$\varepsilon<1$ and $(\Lambda_\text{eff}/M_p^{\,p})_\text{lumpy}<(\Lambda_\text{eff}/M_p^{\,p})_\text{symmetric}$].  The opposite is also true: Whenever the $\ell=2$ `danger mode' has a positive mass squared, the ellipsoidal solution is football-shaped and energetically disfavored [$\varepsilon>1$ and $(\Lambda_\text{eff}/M_p^{\,p})_\text{symmetric}<(\Lambda_\text{eff}/M_p^{\,p})_\text{lumpy}$].
\item The $n\rightarrow0$ behavior is independent of the value of $\Lambda_D$.
\end{itemize}

\subsection{Ellipsoidal Solutions: the Case $\bm{p=q=4}$ and $\bm{\Lambda_D>0}$}
Let us first look in detail at the special case of compactifications of 8-dimensional de Sitter space down to 4 dimensions, where the internal manifold is topologically a 4-sphere.  This is an interesting case study because, as we saw in the Sec.~2.2, it features a phase transition in the Freund-Rubin solutions from stability to instability.

\begin{figure}[t] 
   \centering
   \includegraphics[width=\textwidth]{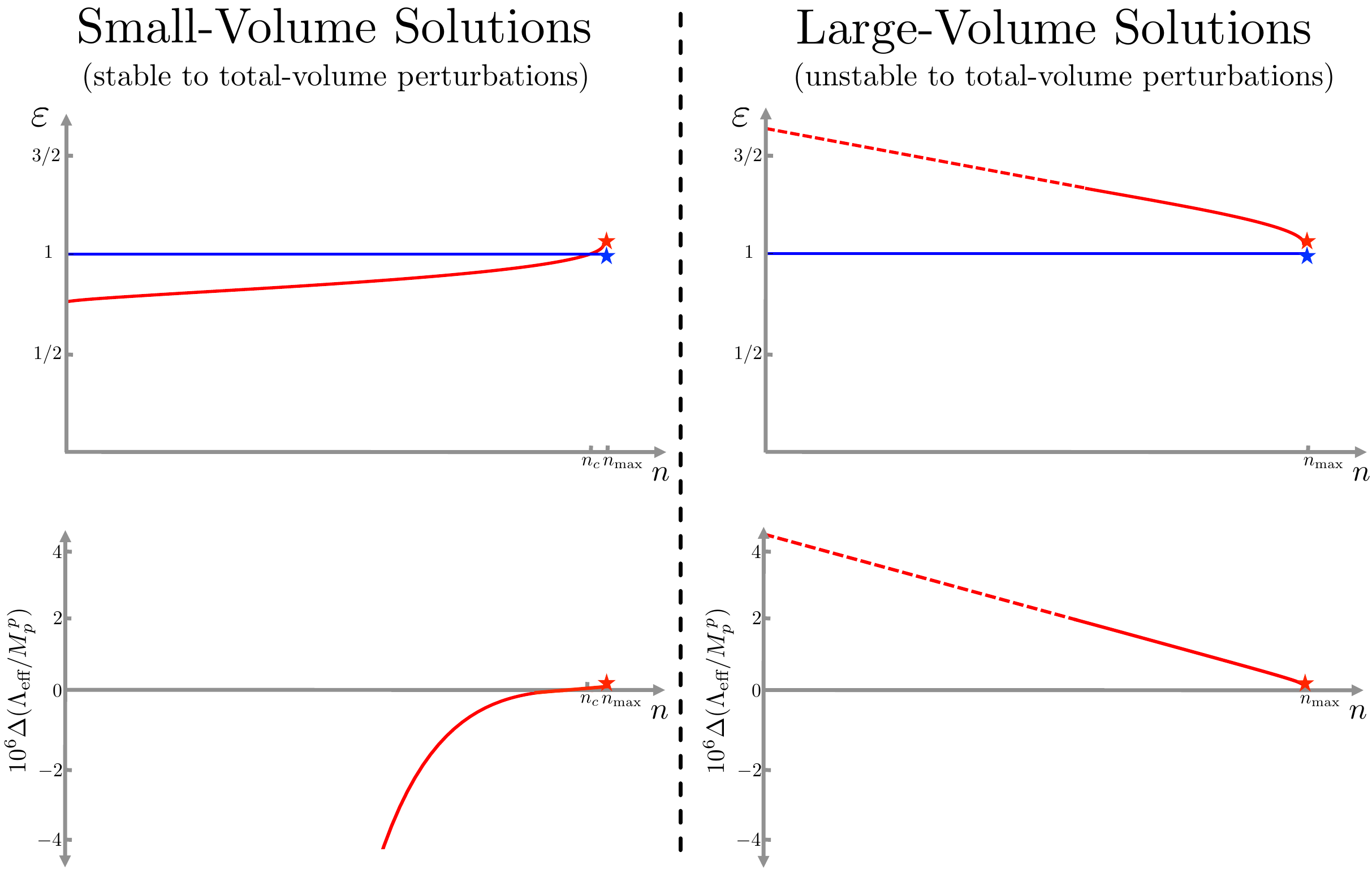}
   \caption{A phase diagram of the ellipsoidal solutions for the case $p=q=4$ and $\Lambda_D>0$.  Symmetric solutions have $\varepsilon=1$; for $n<n_\text{max}$ there are two symmetric solutions---a small-volume solution and a large-volume solution---and for $n>n_\text{max}$ there are no symmetric solutions.   We find that each symmetric solution is accompanied by an ellipsoidal solution with $\varepsilon\neq1$, which has roughly the same internal volume.   Whenever the ellipsoidal solution has $\varepsilon>1$, it also has $\Delta(\Lambda_\text{eff}/M_p^{\,p})\equiv(\Lambda_\text{eff}/M_p^{\,p})_\text{lumpy}-(\Lambda_\text{eff}/M_p^{\,p})_\text{symmetric}>0$ and vice versa.  To produce this phase diagram, we numerically found lumpy solutions at a large number of values of $Q$, and then numerically integrated each solution to find $n$ and $\Lambda_\text{eff}$.  (The region with very small $Q$ is hard to access numerically; this region corresponds to the large-volume ellipsoidal solutions with small $n$.  We were not able to find solutions in this region, and so we have drawn a  straight dashed line that continues the trend.)  
   }
      \label{fig:8Ddsphasediagram}
\end{figure}

As a reminder, there are two critical values of the flux number $n$ that arise  in this case: $n_\text{max}$ and $n_c\sim.97n_\text{max}$.
When $n<n_c$, the small-volume branch is unstable to becoming ellipsoidal while the large-volume branch is stable to it; when $n_c<n<n_\text{max}$,  both branches are stable to it; when $n>n_\text{max}$, there are no Freund-Rubin solutions.

The results of our analysis are given in Fig.~\ref{fig:8Ddsphasediagram}.  We find that there are four solutions when $n<n_\text{max}$ and zero solutions when $n>n_\text{max}$; both the small-volume and the large-volume Freund-Rubin vacua are accompanied by their own lumpy branch of solutions.   The large-volume lumpy branch is always football-shaped ($\varepsilon>1$) and always has a larger value of $\Lambda_\text{eff}/M_p^{\,p}$.  The shape of the small-volume lumpy branch depends on $n$.  As anticipated, it crosses through the Freund-Rubin solution at $n=n_c$.  When $n<n_c$, the small-volume lumpy solution is M\&M-shaped and has a smaller value of $\Lambda_\text{eff}/M_p^{\,p}$, and when $n_c<n<n_\text{max}$, the small-volume lumpy solution is football-shaped and has a larger value of $\Lambda_\text{eff}/M_p^{\,p}$.

These results are consistent with the three principles listed at the start of this section.

\subsection{Ellipsoidal Solutions: the General Case}
\begin{figure}[t] 
   \centering
   \includegraphics[width=\textwidth]{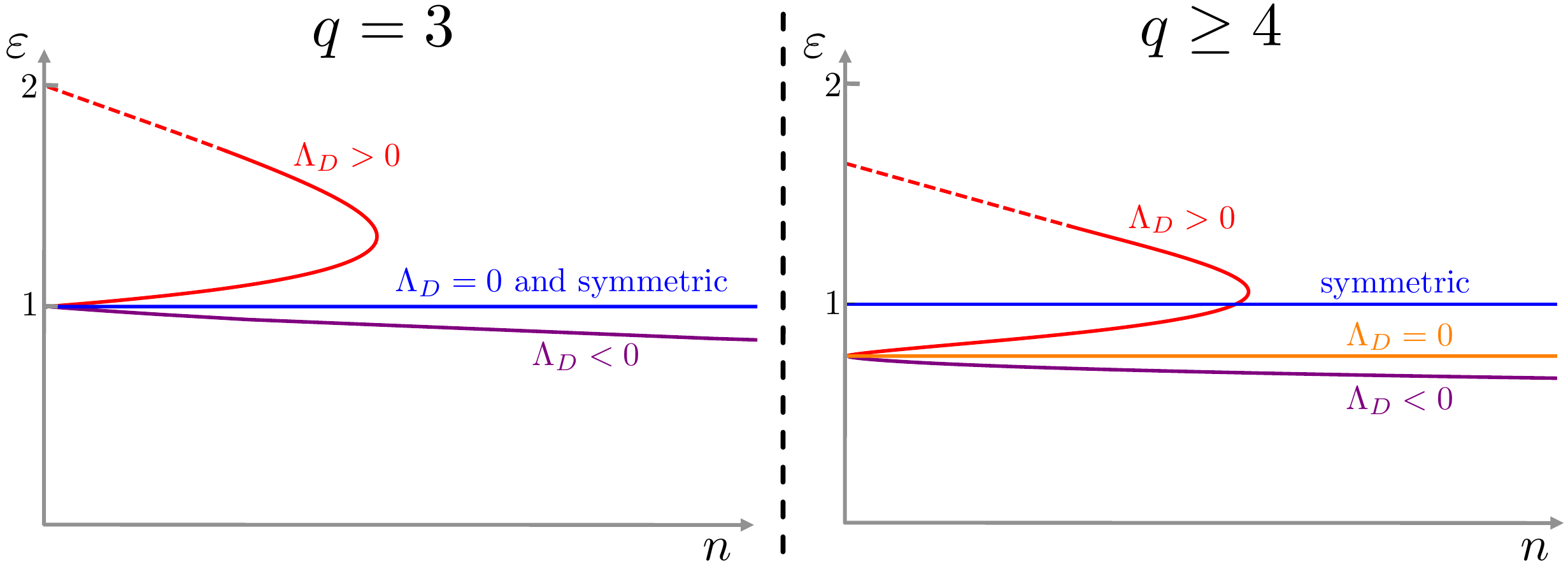}
   \caption{The phase diagram of ellipsoidal solutions.  In all cases, solutions have $\varepsilon<1$ if and only if $\Delta(\Lambda_\text{eff}/M_p^{\,p})<0$ and they have $\varepsilon>1$ if and only if $\Delta(\Lambda_\text{eff}/M_p^{\,p})>0$.  When $\Lambda_D<0$, increasing $n$ makes the solution more and more M\&M-shaped; when $\Lambda_D=0$ the additional scaling symmetry means that $\varepsilon$ stays constant with $n$ and only the total internal volume changes; when $\Lambda_D>0$, there are two ellipsoidal solutions that come together, merge, and annihilate as $n$ is increased. 
   The case $q=3$ is special because there is no ellipsoidal solution when $\Lambda_D=0$; this is related to the fact that the $\ell=2$ `danger mode' is precisely massless.  For all higher $q$, the $\Lambda_D=0$ ellipsoidal solution is M\&M-shaped.
   As $n\rightarrow0$, all three cases merge to a single solution whose volume and shape is independent of $\Lambda_D$.  To make this figure, we numerically found solutions for a large number of values of $Q$ with  $q=3$, 4 and 5, and with $\Lambda_D<0$, $\Lambda_D=0$ and $\Lambda_D>0$.  The specific plots shown here are for $p=4$ and $q=3$ on the left and $p=4$ and $q=4$ on the right.  We conjecture that this qualitative behavior continues for $q\ge4$.  (As with Fig.~8, numerics prevented us from finding solutions with very small $Q$ and $\Lambda_D>0$, and so as before we have drawn a straight dashed line that continues the trend.  Also, in the $\Lambda_D>0$ case, the symmetric branch ends spontaneously at the same $n=n_\text{max}$ as the lumpy branch.) 
   }
      \label{fig:generalqbehavior}
\end{figure}

Figure~\ref{fig:generalqbehavior} represents the phase diagram of ellipsoidal solutions in a wider number of cases, including different numbers of internal dimensions and different signs of $\Lambda_D$.  In all these cases, the three principles listed at the start of this section operate: there is a single ellipsoidal solution for each Freund-Rubin solution, and it has $\varepsilon<1$ if and only if $\Delta(\Lambda_\text{eff}/M_p^{\,p})<0$, and vice versa.

The three cases $\Lambda_D>0$, $\Lambda_D=0$ and $\Lambda_D<0$ behave qualitatively differently.  When $\Lambda_D>0$, there are two branches of lumpy solutions---a small-volume branch and large-volume branch---that come together, merge, and annihilate as $n$ is increased; this is the behavior seen in the explicit example of the previous subsection.  When $\Lambda_D=0$, there is a single ellipsoidal solution for each value of $n$, and its ellipticity is constant in $n$.  Increasing $n$ increases the internal volume, but does not change the shape.  This is related to the extra scaling symmetry of the $\Lambda_D=0$ version of the equations of motion, as discussed in the previous section.
Finally, when $\Lambda_D<0$, increasing $n$ makes the solution more and more M\&M-shaped.

As $n\rightarrow0$, the ellipsoidal solutions converge to a single value $\varepsilon$, independent of $\Lambda_D$.  This is consistent with the fact, discussed in Sec.~2, that when $n$ is very small, the solution has a very small internal volume and there is a large separation of scales between the compactification scale and the higher-dimensional Hubble scale.

\newpage 
The case $q=3$ is qualitatively different from the case $q\ge4$.  When $q=3$ and $\Lambda_D=0$, we find no ellipsoidal solution, whereas for all larger values of $q$, there is an ellipsoidal solution and it is M\&M-shaped.  The reason for this absence can be traced back to the mass squared of the $\ell=2$ `danger mode': when $q=3$, this deformation is exactly massless for $\Lambda_D=0$, and so we find no ellipsoid. However, when $q\ge4$, this deformation has negative mass squared for $\Lambda_D=0$, and so we find an M\&M-shaped solution. 

Once again, these results are all consistent with our three principles.

\section{Higher-$\ell$ Solutions}
\label{higherell}

In the previous section, we numerically studied the lumpy solutions with a single extremum between $\theta=0$ and $\theta=\pi$.  These solutions are  ellipsoidal in shape and are associated with the $\ell=2$  spherical harmonic.  In this section, we report on the behavior of lumpier solutions with additional extrema between $\theta=0$ and $\theta=\pi$.  A sample lumpier solution is given in the bottom row of Fig.~\ref{fig:samplesols}.  These solutions with extra extrema were first found in \cite{Lim:2012gh}; they have much in common with the oscillating bounces of \cite{Hackworth:2004xb,Batra:2006rz,Brown:2011um,Battarra:2013rba}.  (Our assumption of $\mathbb Z_2$ symmetry relating the north and south hemispheres limits our study to solutions with an odd number of extrema.)

\begin{figure}[t] 
   \centering
   \includegraphics[width=\textwidth]{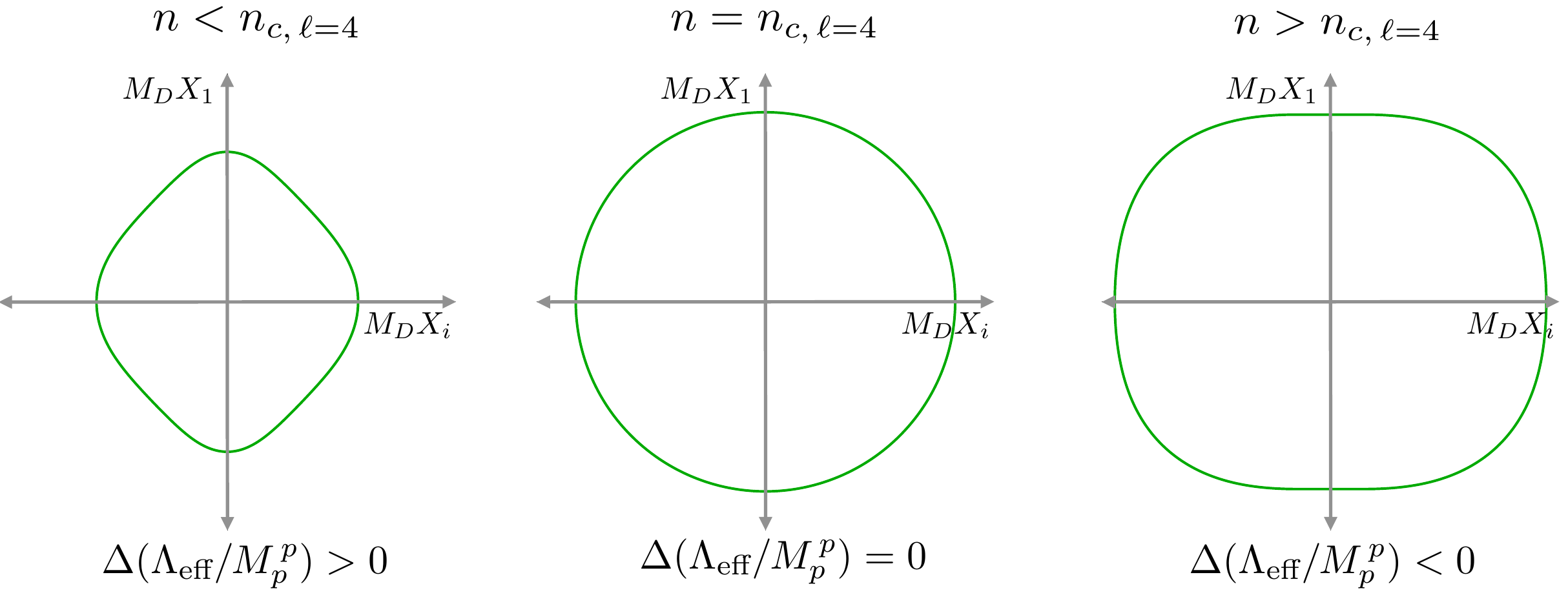}
   \caption{A branch of lumpy solutions associated with $\ell=4$ deformations of the Freund-Rubin solutions; the plots of $R(\theta)$ and $\Phi(\theta)$ for these solutions have three extrema.  The geometry of the internal manifold is shown in embedding coordinates: rotating the surface around the $X_1$-axis gives a surface with an induced metric that matches the lumpy internal metric.  These solutions are for the case $p=q=4$ and $\Lambda_D<0$.  This case has a critical value of $n=n_{c,\,\ell=4}$ at which the $\ell=4$ danger mode develops a negative mass squared.  When $n<n_{c,\,\ell=4}$, the solutions we find always are energetically subdominant and diamond-shaped; when $n>n_{c,\,\ell=4}$, the solutions are energetically favored and box-shaped.  This type of transition is qualitatively similar to that observed for the ellipsoids of the previous section.}
      \label{fig:ell4embedding}
\end{figure}

As before, let us begin with a special case.  Consider compactifying 8-dimensional AdS ($\Lambda_8<0$) down to 4 dimensions.  This case features a phase transition where the $\ell=4$ mode develops a negative mass squared.  (In fact, increasing $n$ eventually makes perturbations with arbitrarily high $\ell$ develop negative mass squareds).  The critical value of $n$ in this case is $n_{c, \,\ell=4}=10\sqrt{330}\pi^2/|\Lambda_8/M_8^8|^{3/2}$.  When $n<n_{c, \,\ell=4}$, the $\ell=4$ perturbation has positive mass squared and when $n>n_c$, it has negative mass squared.  Figure~\ref{fig:ell4embedding} shows a lumpy solution associated with the $\ell=4$ instability for $n<n_{c, \,\ell=4}$, $n=n_{c, \,\ell=4}$ and $n>n_{c, \,\ell=4}$.  As with the ellipsoidal solutions, we find a connection between the shape of the solution and whether it is dominant or subdominant to the Freund-Rubin solution.  In particular, we find solutions that are diamond-shaped and subdominant for small $n$ and box-shaped and dominant for large $n$.

Our investigation of these lumpier solutions was less extensive than for the ellipsoidal solutions; we constructed solutions with one, three, and five extrema for a variety of values of $q$ and $\Lambda_D$.  The behavior of these solutions always exhibits the same connection, which is the reason we conjecture that:
\begin{itemize}
\item Each Freund-Rubin solution is accompanied by a single solution with $k$ extrema for all $k$.
\item These lumpy solutions are related to fluctuations with $\ell=k+1$.  When the $\ell$th danger mode has a negative mass squared, the lumpy solution is dominant and perturbed in one direction along the spherical harmonic $Y_\ell(\theta)$.  When the danger mode has a positive mass squared, the lumpy solution is subdominant and perturbed in the other direction.
\end{itemize}
A more thorough treatment of these lumpier solutions would be interesting, in order to further  test these principles.

\section{The Effective Potential for Lumpiness}

We have just constructed a phase diagram of solutions to Eq.~\eqref{0Einstein}; we found many branches of lumpy solutions that cross the Freund-Rubin solution.  Despite this complexity, we argued that the important features are controlled by simple rules: the classical and thermodynamic stability of the Freund-Rubin solutions determines the shape and behavior of the lumpy solutions.

We will now give an effective potential description that neatly encapsulates these rules.  In Sec.~2.1 we reviewed a $p$-dimensional effective theory that is a dimensional reduction of Eq.~\eqref{action}; we treated the shape of the internal manifold as a rigid sphere and its  radius as a dynamical radion field.  In the effective theory, the radion lives in the effective potential of Eq.~\eqref{effpotjustR}.    Extrema correspond to Freund-Rubin vacua, and the value of the potential at an extremum is $\Lambda_\text{eff}$.  However, we have seen that the shape of the internal manifold does not always remain rigid: we must allow it to vary as well.  In addition to the radion, we must treat the ellipticity $\varepsilon$ as a $p$-dimensional field and extend the effective potential in the $\varepsilon$-direction.  (For the $\ell>2$ modes, we should parametrize lumpiness by other order parameters and extend the effective potential in those new directions as well.)

Because the symmetric Freund-Rubin compactifications are solutions to the equations of motion, the effective potential necessarily has an extremum at $\varepsilon=1$, so that $\partial_\varepsilon V_\text{eff}(\varepsilon=1)=0$.  Whether the extremum is a local minimum or maximum follows from the calculation of the mass squared in Sec.~2.2.  In fact, we saw that varying $n$ can bring about a phase transition---the Freund-Rubin solution can change from a local minimum in the $\varepsilon$-direction to a local maximum in the $\varepsilon$-direction.  

In the spirit of Landau, to understand this phase transition, we look to the next higher-order term in the effective potential: the $(\varepsilon-1)^3$ term.  The presence or absence of locally cubic behavior in the potential dictates the structure of the phase transition.  The potential we sketched in Fig.~\ref{fig:effpotcartoon} exhibits such behavior, and we have seen repeatedly  that the resulting phase transition fully captures the physics of the phase diagrams we constructed numerically.  Had there been no cubic behavior, the phase transition would have proceeded qualitatively differently: two lumpy branches would have merged and annihilated at the symmetric solution when $n=n_c$. The cubic behavior determines the physics.

Because prolate spheroids are different from oblate spheroids, there is no symmetry forcing $V_\text{eff}$ to be even about $\varepsilon=1$ and no reason for a cubic term to be absent.  In fact, it can be computed explicitly in the regime near $n\sim n_c$, where the quadratic term is tuned to be small and therefore the coupling between the $\ell=2$ mode and the higher-$\ell$ modes is also small.  This technique is analogous to that used by Gubser~\cite{Gubser:2001ac} to study the lumpy black strings related to the onset of the Gregory-Laflamme instability~\cite{Gregory:1993vy}.

So far, we have been considering the effective potential in a Taylor series about $\epsilon=1$.  We will now speculate about what happens further from sphericality.  There are three possibilities: first, the effective potential could continue its trend, asymptoting to $V\rightarrow+\infty$ in the infinite M\&M-direction and to $V\rightarrow-\infty$ in the infinite football-direction; second, it could turn over in one or both directions, introducing more extrema; third, it could asymptote to a finite value.  However, scanning over $\Phi(\pi/2)$ did not reveal any such additional solutions, which we consider an argument against option two.
Also, option three implies that the potential has a long flat section without an obvious symmetry protecting it, which we find problematic.  This leaves option one.  

Moreover, we can give an intuitive argument for option one.  In the direction of increasing $\varepsilon$, the $(q-1)$-sphere at the equator is shrinking to zero radius while the flux is clearing away and concentrating at the poles.  Without flux to buttress it, the $(q-1)$-sphere's tendency is to collapse to zero radius; the sphere's curvature term drives the effective potential to $-\infty$ as $R\rightarrow0$.  (This can be seen from the $n=0$ version of Eq.~\eqref{effpotjustR}.)  On the other hand, in the direction of decreasing $\varepsilon$, the $(q-1)$-sphere around the equator is growing to larger size, and the flux is concentrating there.  The north and south poles are moving towards each other in embedding space, but this does not contribute a curvature term to the effective potential.  This argument  relies on the curvature term of the equatorial $(q-1)$ sphere; when $q=2$, this term is 0, and indeed we found no lumpy solutions in that case.

\subsection{Tunneling in the Football-Direction}
Not only does the effective potential explain the details of the phase transition, it also predicts a non-perturbative instability for the perturbatively stable Freund-Rubin vacua with $n>n_c$.  The minimum in the right panel of Fig.~\ref{fig:effpotcartoon} is unstable to the quantum nucleation of bubbles, where the inside of the bubble has football-shaped internal geometry, and the outside has spherical internal geometry. These Freund-Rubin vacua were already known to have two other non-perturbative instabilities: decompactification, in which the internal manifold swells over the potential barrier in Fig.~4, and flux tunneling, in which flux discharges through a Schwinger process.  A rough estimate suggests that shape-mode tunneling proceeds exponentially fastest for all but the highest de Sitter vacua, where decompactification is so fast that it is no longer semiclassical.
\newpage
The fact that shape-mode tunneling typically proceeds faster than decompactification can be seen from the scale of the potential: the effective potential in the shape-mode direction is far weaker than it is in the total-volume direction.  This manifests in two ways.  First, characteristic values of $(\Lambda_\text{eff}/M_p^{\,p})_\text{lumpy}-(\Lambda_\text{eff}/M_p^{\,p})_\text{small-vol FR}$ are about two orders of magnitude smaller than  values of $(\Lambda_\text{eff}/M_p^{\,p})_\text{large-vol FR}-(\Lambda_\text{eff}/M_p^{\,p})_\text{small-vol FR}$.  Second, typical mass squareds in the shape direction are also about two orders of magnitude smaller than mass squareds in the total-volume direction.   The fact that the scale of the potential is smaller suggests that shape-mode tunneling will typically beat volume-mode tunneling.  We can make this statement more quantitative by using the Hawking-Moss instanton to estimate the decay rate \cite{Hawking:1981fz}.  In the Hawking-Moss process, the instanton perches uniformly on the saddle point that separates the true and false vacua, and the decay rate is given by
\begin{gather}
\Gamma\sim e^{-\Delta S_\text{E}}~,\hspace{.2in} \Delta S_\text{E} = S_{\text{E, saddle}}-S_{\text{E, instanton}}~, \hspace{.2in}\text{       where       } \hspace{.2in}S_\text{E}\sim-\left(\frac1{\Lambda_\text{eff}/M_p^{\,p}}\right)^{(p-2)/2}~.
\end{gather}
This gives an estimate of the true decay rate that is increasingly accurate for higher de Sitter vacua, which are also the few that are perturbatively stable to begin with.  For the case $p=q=4$ and $\Lambda_D>0$, we compared Hawking-Moss decay rates and found that 99\% of perturbatively stable vacua prefer to decay by shape-mode tunneling over decompactification.  The rate of flux tunneling depends on the mass of the brane that discharges the flux.  If this mass is tuned to be very small, the rate of flux tunneling can be made arbitrarily fast.  However, for extremal branes whose mass is set by their charge, the no-backreaction estimate of the flux tunneling shows that it proceeds even slower than decompactification.  Shape-mode tunneling is typically the fastest decay.  (For those keeping track, this means that, for $p=q=4$ and $\Lambda_D>0$, 97\% of the Freund-Rubin vacua are perturbatively unstable, 2.97\% decay to footballs, and .03\% decompactify non-semiclassically; the Freund-Rubin branch is decimated.)

\section{Discussion}

The lumpiness instability of Freund-Rubin solutions has a lot in common with the Jeans instability of uniform dust: it is a classical, symmetry-breaking instability in which energy density concentrates in regions of stronger gravity.  (Other analogous examples include the Gregory-Laflamme instability of \cite{Gregory:1993vy} and the striped-phase instability of \cite{Donos:2013gda,Hartnoll:2014gaa}.)  The Jeans instability is ultimately cut off by non-linear terms---collapsing dust forms stars.  Does something similar happen for the classical instability of the Freund-Rubin solution?

We have investigated perturbations that break the SO$(q+1)$ symmetry of Freund-Rubin down to SO$(q)\times\mathbb Z_2$.  Within the dominion of this symmetry, this instability can lead in one of two directions: the flux can either concentrate at points at the poles of the solution, or it can concentrate in bands along lines of constant latitude.  (M\&M-shaped solutions, for example, have a band of flux around the equator and football-shaped solutions have points of flux at the poles.  Higher-$\ell$ solutions combine these features; for instance the $\ell=4$ solution in the left-most panel of Fig.~10 has flux concentrated both at the poles and in a band around the equator.)  We will discuss in turn both possibilities, as parametrized by whether $\epsilon$ gets bigger or smaller from $1$.

When the Freund-Rubin solution is perturbed to smaller $\varepsilon$ in the left panel of Fig.~\ref{fig:effpotcartoon}, the flux starts to concentrate in a band.  In this case, we found a lumpy endpoint: the M\&M-shaped solution.  Consistent with the correlated stability conjecture \cite{Gubser:2000mm}, classical instability is accompanied by thermodynamic instability, and the M\&M's, when they exist, always have $\Delta(\Lambda_\text{eff}/M_p^{\,p})<0.$  However, they are not necessarily the final endpoint of the instability; they might be saddle points with further instabilities that lead to the true endpoint.  Indeed, taking the analogy with the Jeans instability seriously suggests that the M\&M solutions we found, and indeed all the higher-$\ell$ solutions with bands of flux, are likely unstable to perturbations that further break the internal symmetry---if the energy density is trying to clump, it won't be satisfied by a uniform band, it will want to concentrate to a point.  (Specifically, while the mechanism we've detailed in this paper explicitly stabilizes the harmonics with zero azimuthal part, the harmonics with non-zero azimuthal part, which break the remaining symmetry, could still be unstable.)

When the Freund-Rubin solution is perturbed to larger $\varepsilon$ in the left panel of Fig.~\ref{fig:effpotcartoon}, the flux starts to concentrate at the tips.  In this case, we did not find an endpoint; instead, we argued that the effective potential continues its downward trend forever.
What happens to the solution as it rolls down the effective potential, becoming increasingly football-shaped, is unclear; we highlight two possibilities.  The first is analogous to the endpoint of the Jeans instability: as the flux concentrates at the poles,  it can collapse to form a soliton supported by non-linear terms in the potential---such solutions might resemble the rugby-ball solutions of \cite{Aghababaie:2003wz,Burgess:2011va}.  The second is more radical: as the flux concentrates at the poles, the equatorial $(q-1)$-sphere is unsupported by flux; when a sphere is unbuttressed, it can shrink to zero size in finite time, pinching off in a process described in \cite{Adams:2005rb,Brown:2011gt}.
Perhaps becoming football-shaped is the first step towards the sphere ripping itself in two.

The perfect symmetry of the Freund-Rubin solutions is not enough to ensure their stability.  Unstable shape-mode perturbations break the internal symmetry spontaneously.  An important question is to what extent these Freund-Rubin solutions serve as toy models for the far more complex string compactifications. For vacua not protected by supersymmetry~\cite{Cvetic:1992st, PhysRevLett.48.1776,Witten:1981mf, Gibbons:1983aq, 2004CMaPh.244..335D, Hertog:2003ru}, one may worry that the lesson here carries over: symmetries may be broken, lumps may grow. 

\subsection*{Acknowledgements}
We thank Adam R.~Brown, Cliff Burgess, Brian Henning, Kurt Hinterbichler, Matthew C.~Johnson, Janna Levin, and David J.~E.~Marsh. The work of CZ is supported by the Berkeley Center for Theoretical Physics, by the National Science Foundation (award numbers 1214644 and 1316783), by the Foundational Questions Institute grant FQXi-RFP3-1323, by ÒNew Frontiers in Astronomy and CosmologyÓ, by the U.S. Department of Energy under Contract DE-AC02-05CH11231, and by an NSF Graduate Fellowship.

\bibliographystyle{utphys}
\bibliography{mybib.bib}

\end{document}